\documentclass[aps,prb,twocolumn,showpacs,showkeys,footinbib,superscriptaddress]{revtex4-2}

\usepackage{amsmath}
\usepackage{amssymb}
\usepackage{graphicx}
\usepackage{bm}
\usepackage{color}
\usepackage{relsize}
\usepackage{braket}
\usepackage[caption=false]{subfig}
\usepackage{hyperref}
\usepackage[all]{hypcap}
\usepackage{multirow}

\begin{document}
\title{Energy renormalizations of resident carriers and excitons in transition metal dichalcogenide monolayers}
\date{\today}

\author{Dinh Van Tuan}
\email[]{vdinh@ur.rochester.edu}
\affiliation{Department of Electrical and Computer Engineering, University of Rochester, Rochester, New York 14627, USA}
\author{Junghwan Kim}
\affiliation{Department of Electrical and Computer Engineering, University of Rochester, Rochester, New York 14627, USA}
\author{Hanan~Dery}
\email[]{hdery@ur.rochester.edu}
\affiliation{Department of Electrical and Computer Engineering, University of Rochester, Rochester, New York 14627, USA}
\affiliation{Department of Physics and Astronomy, University of Rochester, Rochester, New York 14627, USA}

\begin{abstract} 
Energy renormalizations of resident carriers and excitons are studied theoretically, and compared with recent experiments of electrostatically-doped WSe$_2$ monolayers.  The calculated energy renormalization of resident carriers, subjected to strong out-of-plane magnetic field, reveals the importance of dynamical screening in transition metal dichalcogenides. The  energy renormalization of tightly bound excitons is analyzed through the exchange interaction between the electron (or hole) component of the exciton and resident carriers that share the same spin and valley quantum numbers. Our theory explains the weak energy shift of excitonic resonances despite the strong energy renormalization of resident carriers. We identify the dependence of the energy renormalization on the envelope function of a tightly-bound exciton, showing that unlike free electron-hole pairs, this energy renormalization is not the added renormalizations of a resident electron and resident hole.
\end{abstract}

\pacs{}
\keywords{}

\maketitle
\section{Introduction}

Transition metal dichalcogenides (TMDs) are ideal testbed to study many-particle physics in two-dimensional (2D) systems.  When embedded between materials with weak dielectric screening, the atomically thin nature of TMDs renders a strong Coulomb interaction between conduction band (CB) electrons or valence band (VB) holes \cite{Wang_RMP18,Cudazzo_PRB11,Meckbach_PRB18,VanTuan_PRB18,Marauhn_PRB23}. We refer to these electrons or holes as resident carriers whose density is controlled by electrostatic gating. In TMD moir\'{e} superlattices, the strong Coulomb interaction between resident carriers facilitates strongly correlated phases such as Mott insulators, Wigner crystals, unconventional superconductivity, and topological states   \cite{Nuckolls_NatRevMat24,Andrei_NatRevMat21,Du_Science23,Paik_AOP24,Ju_NatRevMater24,Abajo_ACS25,Bernevig_NatPhys25,Xia_Nat26,Xia_Nat25}.

In TMD monolayers, the strong Coulomb interaction can manifest through distribution of resident carriers in CB or VB valleys \cite{Liu_PRL20,VanTuan_PRL22,Wang_NanoLett17,Dijkstra_NatCom25}. For example, the spin-splitting energy of the CB valleys in WSe$_2$ monolayer is $\sim$12~meV \cite{Ren_PRB23,Jindal_PRB25,Kapuncinski_CP21}, suggesting that filling the top valley should start when the  electron density is $\sim$2$\times$10$^{12}$~cm$^{-2}$ if the Coulomb interactions are weak. However, low-temperature experiments of hBN-encapsulated WSe$_2$ monolayers routinely  show that the top spin-split valleys start to fill when the electron density is $\sim$6$\times$10$^{12}$~cm$^{-2}$ \cite{Wang_NanoLett17,VanTuan_PRL22,Dijkstra_NatCom25}. The discrepancy comes from exchange interactions between resident electrons, resulting in a smaller total energy when the electrons are kept in the bottom valley \cite {Mahan_Book,Giuliani_Vignale_Book,Krishtopenko_JPCM11,Oreszczuk_2DMater13}. Similarly, when subjected to strong out-of-plane magnetic field, the exchange energy lowers the total energy of resident carriers, helping them to sustain complete valley and spin polarization at much larger densities than what one would infer from the g-factor of the energy bands \cite{Back_PRL17,Wang_PRL18,Xu_PRL17,Nedniyom_PRB09,Larentis_PRB18}. 

Another ramification of the strong Coulomb interaction is large binding energy of electron-hole pairs (excitons) \cite{Mak_PRL10,Splendiani_NanoLett10,Chernikov_PRL14,He_PRL14,Goryca_NatCom19}, helping them to withstand the screening from resident carriers \cite{Mak_NatMater13,Scharf_JPCM19}. The tight binding between the CB electron and VB hole can further facilitate a rich variety of excitonic complexes when the electron-hole (e-h) pair binds to resident carriers with different spin and valley quantum numbers \cite{Qiu_PRL13,He_NatComm20, Chen_NatComm18,Ye_NatComm18,Li_NatComm18,Barbone_NatComm18,Liu_PRL20b,Choi_PRB24,Dery_PRX25}. When resident carriers are added to the monolayer, the energy shifts of these excitonic resonances are relatively small compared with the exchange-driven energy renormalization of resident carriers \cite{Dery_PRX25}. The goal of this work is to understand the root cause of this difference, and to separately analyze the exchange-driven energy renormalization of resident carriers and excitons.  Unlike the case of a free e-h pair, wherein the CB electron and VB hole move independently, we show that the confined envelope function of a bound exciton plays a role in setting the exchange interaction of its electron (or hole) component with resident electrons (holes). As a result of its small size and charge neutrality, the energy renormalization of an exciton ends up being much smaller than that of a resident carrier. 
 \begin{figure}[t] 
\centering
\includegraphics[width=8.5cm]{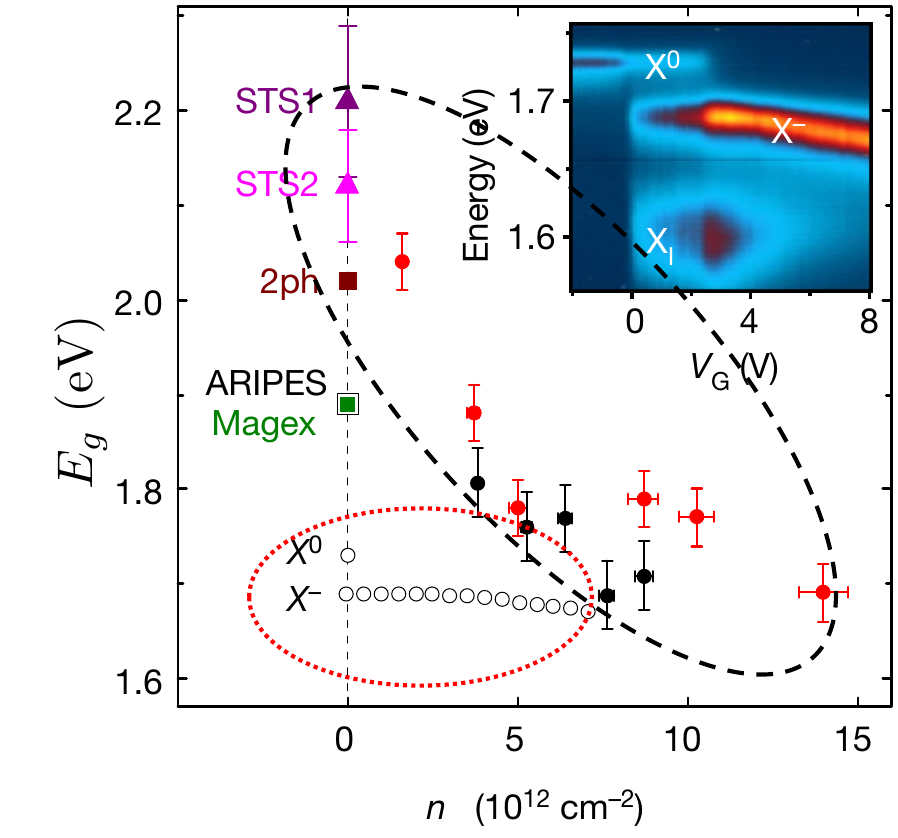}
\caption{  The measured bandgap energy (filled symbols) and excitonic resonance energies (open circles) of WSe$_2$ monolayers as a function of CB electron density. Taken from Ref.
\cite{Nguyen_Nat19}. The inset shows photoluminescence spectra of WSe$_2$ monolayers.   \label{fig:ExpXXu} }
\end{figure}

Before we embark on the theory, we survey three representative experimental results that sharpen the difference between energy renormalizations of resident carriers and excitons.  The first experimental result is shown in Fig.~\ref{fig:ExpXXu}, where angle-resolved photoemission spectroscopy (ARPES) reveals a dramatic bandgap shrinkage when the electrostatic doping is significant (filled symbols in the highlighted dashed oval) \cite{Nguyen_Nat19}. These measurements probe the energy renormalization of CB electrons that are emitted out of the  monolayer. Contrary to this large energy renormalization, the energy shift of the charged exciton resonance is much smaller (open circles in the highlighted dotted oval; $X^-$). 

 \begin{figure}[t] 
\centering
\includegraphics[width=8.5cm]{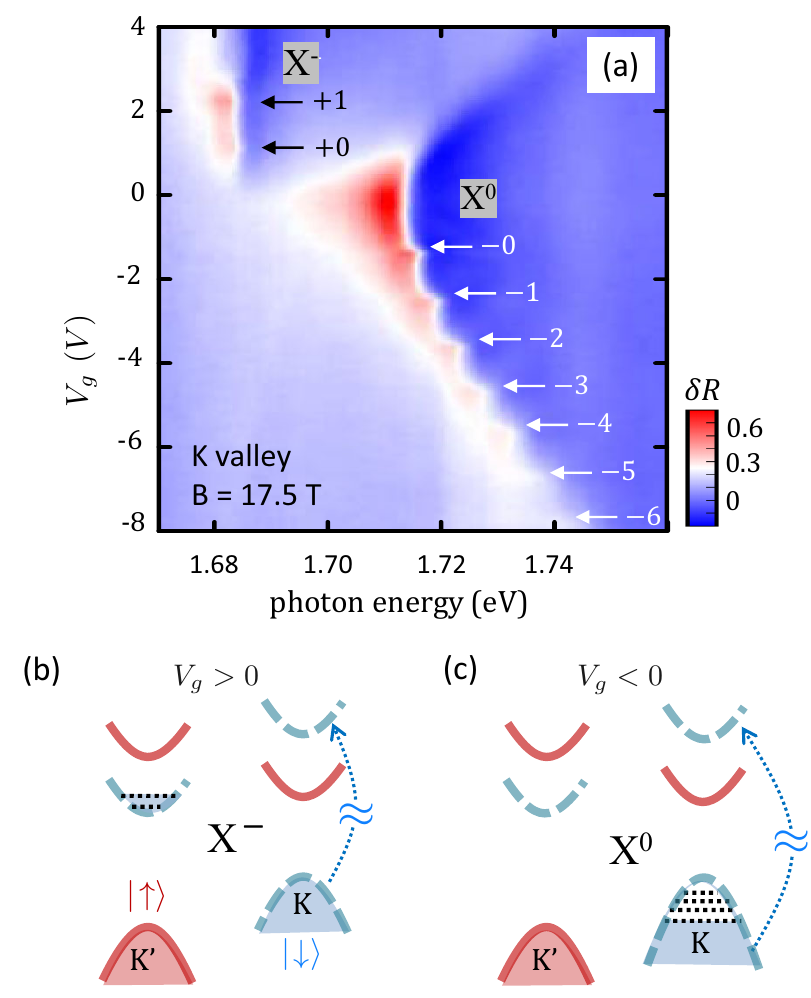}
\caption{  (a) Magneto-optical reflectance spectra from  $K$ valley of a charge tunable WSe$_2$ monolayer at 4 K. Taken from Ref. \cite{Liu_PRL20}. The out-of-plane magnetic field is 17.5 T.  (b) Optical transition of the negative trion when resident electrons only populate the bottom CB valley at $K$'. (c) Optical transition of the exciton when resident holes only populate the VB valley at $K$.}\label{fig:ExpDat2} 
\end{figure}

The second experimental result, taken from Ref.~\cite{Liu_PRL20} and shown in Fig.~\ref{fig:ExpDat2}, is magneto-optical reflectance spectra of the exciton in the $K$ valley of WSe$_2$ monolayer. The magnetic field is $B = 17.5\,$T, and the corresponding degeneracy of each Landau level (LL) is $4.23 \times 10^{11}$~cm$^{-2}$. The presence of LLs can be identified through the staircase shape of the excitonic resonances, applicable when the resident carriers are fully spin and valley polarized  \cite{Dery_PRX25}. Here, seven VB LLs are identified when $V_g<0$, and two CB LLs are identified when $V_g>0$. The window of complete spin and valley polarization of holes is larger because the g-factor of the VB is larger than that of the bottom CB valley ($g_v = 6.1$ vs $g_{c,b} = 0.86$) \cite{Robert_PRL21}. Focusing on the hole-doped regime, the Zeeman energy between the $K$ and $K$' valleys is $\Delta_z = 2\mu_B g_v B \sim 12$~meV. In addition, the energy spacing between LLs in the VB is 5.6~meV at  $B = 17.5\,$T,  suggesting that the Fermi level in the $K$ valley is $E_{F,K}\,$$\sim$$\,$40~meV when seven LLs are filled.   The exchange-driven energy renormalization of holes in the $K$ valley explains the fact that the VB valley at $K$' remains unoccupied even though $E_{F,K} \gg \Delta_z$. 

Viewing the energy renormalization of holes as an effective g-factor enhancement is inconsistent with the notion that the g-factor of an energy band is a single-particle parameter. The underlying effect comes from the exchange component in the total energy of the hole gas. To clarify that the effect does not come from a rigid energy shift of the VB band, we study the energy shift of the exciton resonance $X^0$ in Fig.~\ref{fig:ExpDat2}(a). If the energy renormalization of resident holes led to upward shift of the VB valley at $K$ in Fig.~\ref{fig:ExpDat2}(c), we would expect the exciton resonance to redshift in energy when resident holes with similar spin and valley quantum numbers start to fill the monolayer.  Instead, Fig.~\ref{fig:ExpDat2}(a) shows a steady energy blueshift, reflective of Pauli blocking due to successive LL filling. Therefore, we conclude that energy renormalization of resident carriers has no effective bearing on the energy shift of an exciton even if its electron or hole share similar quantum numbers with the resident carriers.

\begin{figure*}[t] 
\centering
\includegraphics[width=\textwidth]{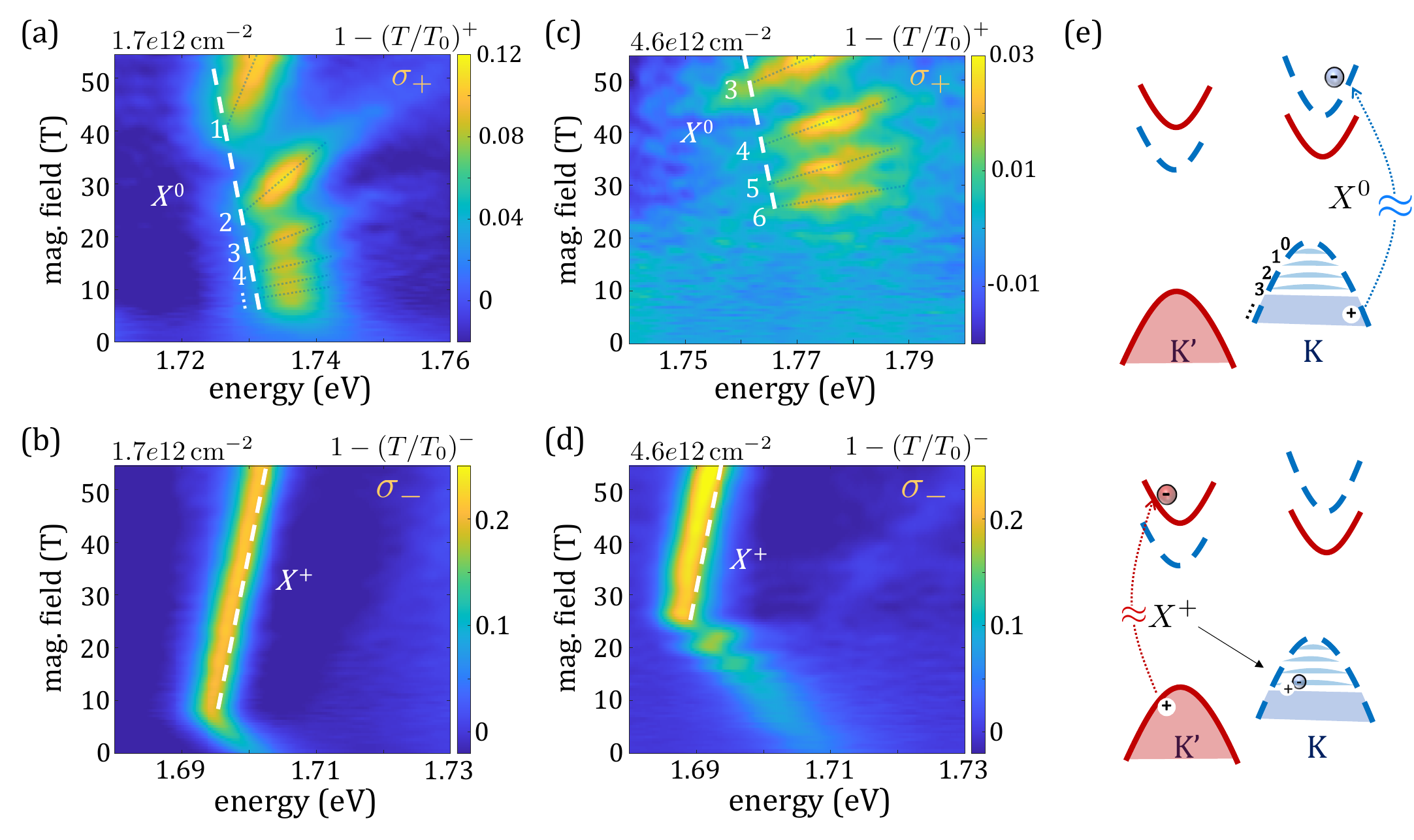}
\caption{  Color maps of  helicity-resolved magneto-optical absorption spectra  of an electrostatically hole-doped WSe$_2$ monolayer as a
function of magnetic field and photon energy at 4 K. Taken from Ref. \cite{Dery_PRX25}. The measured signal is $(1 - T/T_0)^\pm$,  where $T$ $(T_0)$ is the transmission (reference)
spectrum, and $\pm$ refers to light with $\sigma_\pm$ polarization. The hole density is $1.7 \times 10^{12}$ cm$^{-2}$  in (a),(b) and $4.6 \times 10^{12}$ cm$^{-2}$  in (c),(d).
(e) Schemes of optical excitations in valley-polarized WSe$_2$ monolayer with hole LLs in the VB valley at $K$.}\label{fig:Expdat1} 
\end{figure*}

To further reinforce the disconnect between the energy shifts of excitonic resonances and the energy renormalization of resident carriers, we analyze the experimental results of Refs.~\cite{Li_PRL20,Dery_PRX25}. Figure \ref{fig:Expdat1} shows magneto-optical absorption spectra of WSe$_2$ monolayers as a function of magnetic field $B$ when the hole density is $1.7 \times 10^{12}$ cm$^{-2}$  in (a)-(b), and $4.6 \times 10^{12}$ cm$^{-2}$  in (c)-(d) \cite{Li_PRL20}. The upper panel in Fig.~\ref{fig:Expdat1}(e) shows the exciton optical transition, corresponding to the spectra in Figs.~\ref{fig:Expdat1}(a) and (c). Similarly, the lower panel in Fig.~\ref{fig:Expdat1}(e) shows the positive trion optical transition, corresponding to the spectra in Figs.~\ref{fig:Expdat1}(b) and (d). Note that these color maps are $B$-dependent spectra at a fixed carrier density, whereas Fig.~\ref{fig:ExpDat2}(a) shows density-dependent spectra at a fixed magnetic field.

Comparing the field dependence of the exciton and trion resonances reveals interesting behavior. The emergence of exciton resonances in Fig.~\ref{fig:Expdat1}(a) and the onset of energy blueshift of $X^+$ in Fig.~\ref{fig:Expdat1}(b) take place when $B\gtrsim10$~T.  This strong field is needed to fully valley-polarize the monolayer where holes only populate LLs of the VB valley at $K$, as shown in Fig.~\ref{fig:Expdat1}(e). The Fermi energy in the $K$ valley is $E_{F,K}\,$$\sim$$\,$22~meV  when the hole density is $1.7 \times 10^{12}$ cm$^{-2}$, and it is once more about three times larger than the Zeeman energy difference between the $K$ and $K$' in the VB, $\Delta_z \sim 7$~meV, when $B\sim10$~T. In case of Figs.~\ref{fig:Expdat1}(c) and (d), the hole density is 4.6$\times10^{12}$~cm$^{-2}$ and complete valley polarization is achieved when  $B\gtrsim25$~T. The respective parameters are  $E_{F,K}\,$$\sim$$\,$60~meV and $\Delta_z \sim 18$~meV. Despite the fact that $E_{F,K} \gg \Delta_z$ in both cases, we have complete valley polarization because of the exchange interaction between resident holes in the $K$ valley. At the same time, the energy shifts of the positive trion and exciton resonances show no hints of the strong energy renormalization of resident holes. 
 
Starting with the positive trion in Figs.~\ref{fig:Expdat1}(b) and (d), we notice energy blueshift of $X^+$ when $B$ increases. This behavior can be understood from the g-factors of the energy bands \cite{Robert_PRL21}, where the Zeeman energy lowers the VB valley at $K'$ more than it lowers the top spin-split CB valleys at $K'$ (see Fig.~\ref{fig:Expdat1}(e)). The energy blueshift can be modeled through the difference of their g-factors, $g_v-g_{c,t}$, where $g_v \simeq 6.1$ and $g_{c,t} \simeq 3.84$ \cite{Robert_PRL21}. As shown in the lower panel of Fig.~\ref{fig:Expdat1}(e), the positive trion is formed by binding of the photoexcited e-h pair from the valley at $K'$ to a resident hole from the valley at $K$. While the energy of the resident hole is renormalized by the exchange interaction, the trion resonance energy is not affected by this interaction, suggesting that the resident hole does not impart its renormalized energy to the optical process (i.e., it is merely a spectator).

Continuing with the exciton resonance in Figs.~\ref{fig:Expdat1}(a) and (c), we notice a more complex behavior of $X^0$ when $B$ increases. The labeled integers indicate the highest filled LL of the VB valley at $K$. The dotted dark lines trace repeated blueshifts of the exciton resonance, explained by the change in compressibility of resident holes in partially filled LLs \cite{Dery_PRX25}. Here, we focus on the dashed white lines in Figs.~\ref{fig:Expdat1}(a) and (c), which trace the overall energy redshift of $X^0$ at integer filling of the LLs. We identify the role of Pauli blocking in these $B$ dependent color maps by noticing that the exciton resonances in Fig.~\ref{fig:Expdat1}(c) appear above 1.76~eV whereas those in Fig.~\ref{fig:Expdat1}(a) appear below 1.74~eV due to different numbers of filled LLs.  In addition, we note that the slopes of the dashed white lines are the same but opposite in sign to those of $X^+$ in Figs.~\ref{fig:Expdat1}(b) and (d), stemming from the opposite-sign g-factors in opposite valleys, $g_{-K} = -g_{K}$. Namely, the overall energy redshift of $X^0$ reflects the Zeeman energy, where the VB valley at $K$ is raised more than the CB valleys at $K$ (see upper panel of Fig.~\ref{fig:Expdat1}(e)). To see that energy renormalization of resident holes has no bearing on the exciton resonance energy, we point out the following. While Figs.~\ref{fig:Expdat1}(a) and (c) are similar measurements taken at largely different hole densities, the slopes of the dashed white lines are exactly the same in both cases. This result supports the understanding that the energy shift of $X^0$ solely depends on the g-factors of the energy bands. 

The experimental results in Figs.~\ref{fig:ExpXXu}-\ref{fig:Expdat1} raise the following question. Why is the exciton not susceptible to the same energy renormalization of resident carriers even when one of its components (electron or hole) share similar spin and valley quantum numbers. We address this question in this work by separately investigating the  energy renormalization of resident carriers and excitons in TMDs. Our findings identify the importance  of dynamical effects in the self-energy of resident carriers. We show that  the energy renormalization of an exciton is not the combination of self-energies of free-like electron and hole components. We treat the exciton as a bound state, and use its envelop function to derive its energy renormalization due to the interaction with resident carriers. Good agreement is found between the obtained results of the theory and the experimental observations. 

This paper is organized as follows. Section~\ref{sec:BGRe} includes the analysis of energy renormalization of resident carriers under the effect of magnetic field.  We highlight the important contribution of dynamical screening  and compare the calculated results with experimental measurements of electrostatically-doped  WSe$_2$ monolayers.  In Sec.~\ref{sec:BGRX} we develop a theory for the energy renormalization of excitons.  A summary  is  given in Sec.~\ref{sec:sum}. The Appendices include derivations and computational details.

\section{Energy renormalization of resident carriers}\label{sec:BGRe}

We study the energy renormalization of resident carriers in TMDs under the effect of a strong magnetic field.  Electrostatically-doped WSe$_2$ monolayers are selected as a prototype due to the availability of high-quality experimental results \cite{Liu_PRL20,Courtade_PRB17,Nguyen_Nat19,Wang_PRL18,Li_PRL20,Robert_PRL21,Liu_PRL19}. We use the results of Fig.~\ref{fig:ExpDat2}(a) to benchmark our theory, where a magnetic field of $B = 17.5\,$T fully polarizes electrons (holes) up to a density high enough to fill 2 (7) LLs. A complete valley and spin polarization is shown in Figs.~\ref{fig:ExpDat2}(b) and (c) for the electron- and hole-doped regimes, respectively.  We employ the random phase approximation (RPA) to describe the screening caused by resident carriers \cite{Scharf_JPCM19}. By comparing the experimental results with calculations based on the screened Coulomb potential in the static and dynamical regimes, we highlight the important role of  dynamical screening in TMDs. 

To simulate the experiment, we consider $N$ filled LLs, with $N_K$ in the valley at $K$ and $N_{K'}= N-N_{K}$ in the valley at $K$'. The value of $N_K$ is found by minimizing the total  energy of the resident carrier system. The total energy per particle is    \cite{Giuliani_Vignale_Book}
\begin{eqnarray}
\overline{\varepsilon}_T \left(N_K \right) &=& \frac{1}{N} \Bigg[ \sum_{i=0}^{N_K - 1} \left( E_c\left( i\right) +    \frac{1}{2}\Sigma\left( i\right) \right)  - N_K g \mu_\text{B} B  \nonumber\\ 
&\!\!\!+& \!\!\!\!\!\!\!\!   \sum_{i=0}^{N_{K'} - 1} \left( E_c\left( i\right) +   \frac{1}{2} \Sigma'\left( i\right) \right)  +N_{K'} g \mu_\text{B} B \Bigg],\qquad
\end{eqnarray}
where  $\mu_\text{B}$ is the Bohr magneton, and  $g$  is the g-factor of the relevant energy band. The first and second lines correspond to energy contributions from the $K$ and $K'$ valleys, respectively. Each line includes, from left to right, the cyclotron energy of carriers in filled LLs, their self-energy from many-body effects, and Zeeman energy.    The cyclotron energy $E_c\left( i\right) = \hbar \omega_\text{c} \left( i + \frac{1}{2}\right)$ is the kinetic energy of a carrier in the $i^\text{th}$ LL with the cyclotron frequency $\omega_\text{c} =  eB/m$, written in terms of the elementary charge ($e$) and carrier effective mass ($m$). Because the self-energy comes from pair-wise exchange interactions \cite{Peng_RPB25},  the factors $\frac{1}{2}$ in front of $\Sigma(i)$ and $\Sigma'(i)$ prevent double counting.

In the following, we evaluate the self-energy in the static and dynamical screening regimes. In the static limit,  the  self-energy is the screened exchange energy, $\Sigma(i) \equiv \Sigma_\text{sx}(i)$ (see Appendixes \ref{app:SX} and \ref{app:SelfEB})
\begin{equation}
\Sigma_\text{sx}(i) = -\sum_{i'}  f_{i'}   \sum_{\bf q} V_s({\bf q}) \,\,\, \left( \tilde{L}^{|\Delta i|}_{\tilde{i}} \left( x_q \right) \right)^2, 
\label{Eq:SelfSX}
\end{equation}
where $f_{i}$ is the Fermi-Dirac distribution of  carriers in the $i^\text{th}$ LL, and $x_q = \tfrac{1}{2}q^2 \ell_\text{B}^2$ is defined via the  magnetic length $\ell_\text{B} = \sqrt{\hbar/(eB)}$.  The sum over $i'$ represents the exchange interaction of carriers in the  ${i'}^\text{th}$ LL  with a carrier in the $i^\text{th}$ LL of the same valley.    $\Delta i  = i' - i$ and $ \tilde{i}= \text{min}\!\left\{i,i' \right\}$  are the difference and minimum of the two LL indices, respectively. The set of orthonormal functions
\begin{equation}
\tilde{L}^j_i(x) = \sqrt{  \frac{ i! }{    \left( i+ j \right)! } \,\,\,  x^{j} \,\,  e^{-x  }  }       \,\,\,\,     L^{j}_{i}\left( x\right)  \label{Eq:Lmntilde}
\end{equation}
are constructed from the generalized Laguerre polynomials $L^{j}_{i}\left( x\right)$, related to the eigenfunctions of non-interacting carriers in magnetic field \cite{VanTuanPRB25_2}. The statically screened potential in Eq.~(\ref{Eq:SelfSX}) is
\begin{equation}
V_s({\bf q}) \equiv V_s({\bf q},\omega = 0)  = \frac{V({\bf q})}{\epsilon_s({q},\omega = 0)},
\label{Eq:Vs}
\end{equation}
where $V({\bf q})$ is the bare Coulomb potential, and $\epsilon_s({q},\omega = 0)$ is the statically-screened dielectric function 
\begin{equation}
\epsilon_s({q},\omega = 0) = 1 + \frac{AV({\bf q})}{\pi \hbar \omega_c \,\, \ell^2_\text{B}} \left[ F_K({x_q}) + F_{K'}({x_q}) \right] \,.
\label{Eq:es}
\end{equation}
$A$ is the sample area. The valley components of the screening factor $F_K(x_q) + F_{K'}(x_q)$ are calculated from the density-density response function of a 2D electron gas in magnetic field \cite{Giuliani_Vignale_Book}, 
\begin{eqnarray}
F_\alpha(x_q) &=& \sum_{ k = 1}^{\infty}    \frac{1}{k}   \sum_{j=j_\alpha}^{N_\alpha-1} \left( \tilde{L}^{k}_{j} ( x_q ) \right)^2   \!,  \label{Eq:Chi} 
\end{eqnarray}
where $\alpha \!=\! \{ K, K' \}$ and $j_{\alpha} \!=\! \text{max}\{0, N_{\alpha} - k \}$.

\begin{figure}[t] 
\centering
\includegraphics[width=8.5cm]{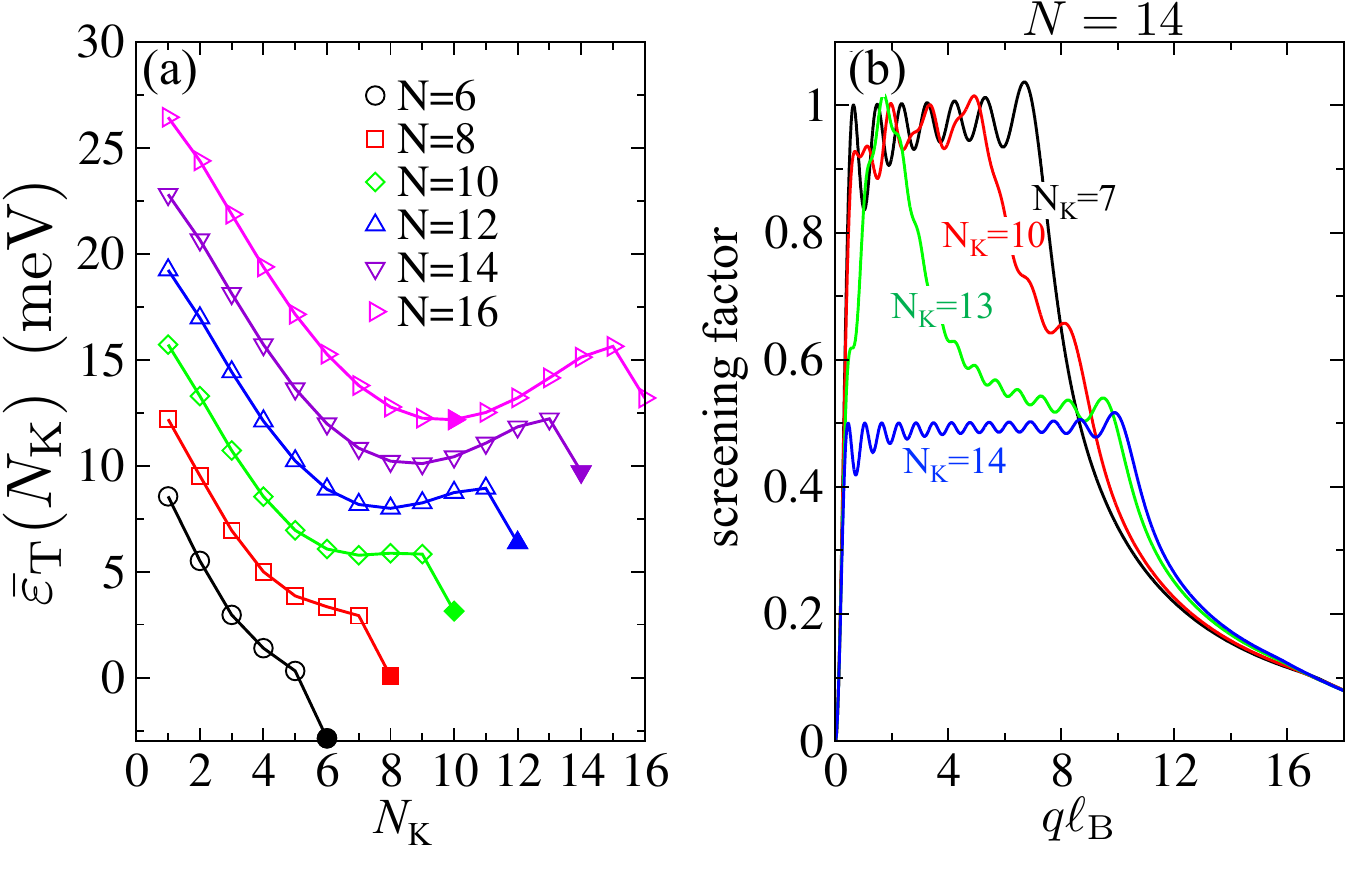}
\caption{ Energy renormalization and screening effects of resident holes  in the static regime. The calculations simulate hBN-encapsulated WSe$_2$ monolayers at $B = 17.5\,$T (see Appendix \ref{app:num} for parameters).  (a) Total energy per particle   as a function of filled LLs in the valley at $K$. The calculations are performed for $N = \{6,8,10,12,14, 16\}$   filled LLs in the two valleys. The  calculated values are marked by open symbols,  among which the minimum values are marked by filled symbols.   (b) The screening factor $F_{K}({x_q}) + F_{K'}({x_q})$ for several cases of $N_K$ when $N=14$. 
 }\label{fig:BGRe} 
\end{figure}

Figure~\ref{fig:BGRe}(a) shows the total energy per particle versus $N_K$ in hole-doped WSe$_2$ monolayer when $B = 17.5\,$T. The corresponding hole density in a filled LL is $4.23 \times 10^{11} $ cm$^{-2}$ \cite{Liu_PRL20}. The calculations are performed for several cases of $N$, where all show initial decrease in the total energy per particle when $N_K$  increases from zero. The initial decrease is explained by the preferred population of the valley at $K$ whose energy is lowered by the Zeeman energy. In addition, when the hole density is less than $\sim$4$\times 10^{12}$~cm$^{-2}$ ($N \lesssim 10$), the total energy per particle decreases continuously until $N_K  = N$ and $N_{K'} = 0$, at which case the hole system is fully valley polarized. At the highest hole density that we study ($N = 16$), the total energy per particle reaches a minimum when $N_K=10$ ($N_{K'} = 6$) LLs are filled in the valley at $K$ ($K$').  In between ($10 <  N \leq 14$), the total energy per particle has a local minimum, which is still greater than the global minimum at full valley polarization. 

\begin{figure*}[t] 
\centering
\includegraphics[width=\textwidth]{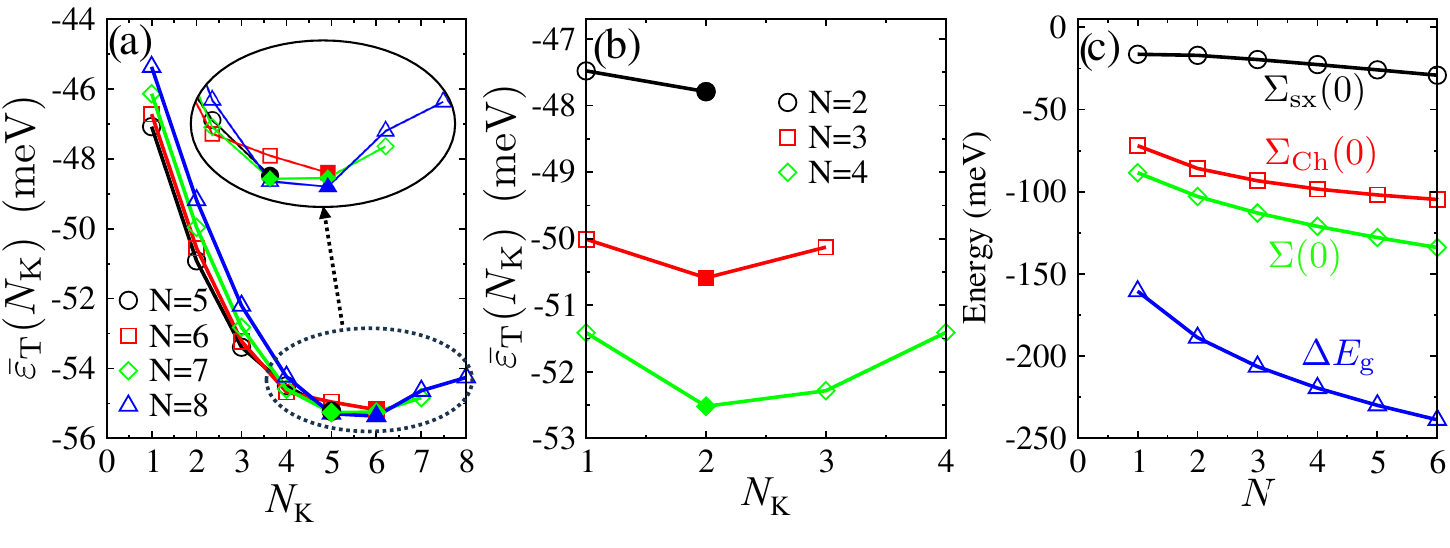}
\caption{ Energy renormalization of WSe$_2$ monolayers in the dynamical regime at $B = 17.5\,$T. (a) Same as in Fig.~\ref{fig:BGRe}(a) but with inclusion of the Coulomb-hole energy.  (b) Same as in (a) but for an electron-doped WSe$_2$ monolayer. (c) Screened exchange and Coulomb-hole energies of the lowest LL in a fully valley-polarized hole system with $N$ filled LLs. The self-energy of holes $\Sigma(0) = \Sigma_\text{sx}(0) + \Sigma_\text{Ch}(0)$ and the bandgap renormalization of free e-h pairs  $\Delta E_\text{g} = \Sigma_\text{sx}(0) + 2 \Sigma_\text{Ch}(0)$ are shown by the green diamond and  blue triangle symbols, respectively. The parameters used in the calculations are detailed in Appendix \ref{app:num}.  } \label{Fig:BothBGR}
\end{figure*}

These results show that holes prefer to populate only one valley up to $N=14$ filled LLs, which is twice the value found in experiment, as shown in Fig.~\ref{fig:ExpDat2}(a) \cite{Liu_PRL20}.  Therefore, calculating the self-energy with the statically screened potential fails to capture the energy renormalization.  As shown in Fig.~\ref{fig:BGRe}(a), complete valley polarization at $N_K = N$ is accompanied by a sharp drop in energy. To understand the reason for this drop, Fig.~\ref{fig:BGRe}(b) shows the screening factor, $F_K({x_q}) + F_{K'}({x_q})$, when $N=14$. In the intermediate range, $ 1 < q\ell_\text{B} <  \sqrt{8(N- N_{K})}$, the static RPA screening approaches the Thomas-Fermi approximation and we get that $F_{K}({x_q}) + F_{K'}({x_q}) \sim 1$. However, the screening is suppressed under complete valley polarization because $F_{K'}({x_q})=0$  when $N_{K'}=0$, resulting in $F_{K}({x_q}) + F_{K'}({x_q}) \sim 1/2$, as shown in Fig.~\ref{fig:BGRe}(b) for $N_K=14$. All in all, the suppressed screening factor when the system is fully valley-polarized contributes significantly  to the self-energy, giving rise to the drop of the total energy per particle when $N_K=N$ in Fig.~\ref{fig:BGRe}(a).

The failure to support the experimental results stems from the omission of dynamical screening (i.e., assuming the static limit $\omega=0$ in Eqs.~(\ref{Eq:SelfSX})- (\ref{Eq:es})). The dynamical effect has been proven to be important in TMD monolayers \cite{Mhenni_ACS25,VanTuanPRB24}, and we include this effect by using the spectral representation of the screened potential. This representation allows us to express the self-energy through the imaginary-time Green function formalism, and then to use analytic continuation from  imaginary Matsubara frequencies to the real energy axis \cite{Haug_SchmittRink_PQE84,Scharf_JPCM19}.  Appendixes  \ref{app:CH} and \ref{app:SelfEB} provide detailed derivations. After  neglecting the recoil energy, the self-energy can be written as %
\begin{equation}
\Sigma(i) = \Sigma_\text{sx}(i)+   \Sigma_\text{Ch}(i)\,,
\label{Eq:totalsig}
\end{equation}
where the first term is the screened exchange energy (Eq.~(\ref{Eq:SelfSX})), and the second term is the dynamically-driven Coulomb-hole energy. The term Coulomb-hole refers to the lack of CB electrons in the vicinity of a CB electron, or equivalently, the lack of VB holes in the vicinity of a VB hole. Its form reads (Appendixes  \ref{app:CH} and \ref{app:SelfEB})   
\begin{equation}
\Sigma_\text{Ch}(i)
=  -\frac{1}{2}    \sum_{i',{\bf q}} \left[ V({\bf q}) \!\!- \!\!V_s({\bf q}) \right] \left( \tilde{L}^{|\Delta i|}_{\tilde{i}} ( x_q ) \right)^2.  
\label{Eq:SelfCh}
\end{equation}
Unlike the screened exchange energy, the Coulomb-hole energy does not require the valley to be populated (no $f_{i'}$ term like the one in Eq.~(\ref{Eq:SelfSX})). It is not the exchange with other particles that introduces the Coulomb-hole renormalization, but the screening-induced change to the Coulomb potential. The change is expressed in Eq.~(\ref{Eq:SelfCh}) by the term in square brackets, and it equally renormalizes the energies of CB electrons and VB holes. This explains the change in bandgap energy of TMD monolayers in different dielectric environments, including in samples without resident carriers \cite{Marauhn_PRB23,VanTuanPRB24,Mhenni_ACS25}.

Figure \ref{Fig:BothBGR}(a)  shows the energy per particle in hole-doped WSe$_2$ monolayer under the same conditions as in Fig.~\ref{fig:BGRe}, but with inclusion of  the Coulomb-hole energy. We no longer observe the sharp energy drop when $N_K = N$, coming from an overall weaker screening effect because of partial cancelation of the screened potential terms in the Coulomb-hole and screened-exchange energies. The result of including the dynamical effect through the Coulomb-hole energy is that VB holes become partially polarized at a lower density compared to the one in the static regime: $N_K = 6$ filled LLs in Fig.~\ref{Fig:BothBGR}(a) compared with evidently larger values of $N_K$  in Fig.~\ref{fig:BGRe}(a).  Another important aspect is the metastability around the minimum configuration. The highlighted region in Fig.~\ref{Fig:BothBGR}(a) shows that the minimum region is relatively flat, with tiny energetic differences between adjacent configurations. For example, the states  $N_K = 5$ and $N_K = 6$ have essentially the same energy when $N=7$ (green diamond symbols). The instability at the verge of transition between complete and nearly complete polarization has been reported experimentally by  Li  \textit{et al.} \cite{Li_PRL20}. Finally, Fig.~\ref{Fig:BothBGR}(b) shows similar calculations carried for an electron-doped WSe$_2$ monolayer at the same magnetic field. The results for $N = \{ 2, 3, 4\}$ show that  the  minimum energy for all three cases is at $N_K =2$. 

All in all, the application of the theoretical model with screened-exchange and Coulomb-hole components yields very good results. Modeling the case of WSe$_2$ monolayers at $B =17.5\,$T, we find that resident holes remain fully polarized up to a density high enough to fill  $N_K=6$ LLs, and that of resident electrons up to $N_K=2$ LLs. The respective reported values in experiment are $N_K=7$ and $N_K=2$ (Fig.~\ref{fig:ExpDat2}).  The slight deviation in the case of resident holes can stem from the instability near the minimum configuration or from overestimation of screening at higher densities due to the RPA. Nonetheless, the RPA is still quite good description of screening effects in TMDs. 

In closing of this section, we compare the contributions from the Coulomb-hole and screened-exchange energies to the renormalization of resident carriers. Figure~\ref{Fig:BothBGR}(c) shows this comparison when $N<7$ in a system of resident holes at $B =17.5\,$T, such that the holes remain fully valley-polarized ($N_K=N$).  The screened exchange energy  of a resident hole in the lowest LL reaches 30~meV at $N = 6$ filled LLs (corresponding to a hole density of $2.5 \times 10^{12}$ cm$^{-2}$), while the corresponding Coulomb-hole energy is over 100~meV.  Namely, the Coulomb-hole energy is the dominant contribution. The green diamond symbols show the sum of screened-exchange and Coulomb-hole energies, which in our case represents the energy renormalization of VB holes in the lowest LL. To obtain the bandgap renormalization of a free e-h pair, $\Delta E_\text{g}$, we should include the energy renormalization of the CB electron. Because the CB band is not populated by resident electrons, the electron component of the free e-h pair only includes the Coulomb-hole energy $\Sigma_{\text{Ch}}(0)$, leading to $\Delta E_\text{g} = \Sigma_{\text{sx}}(0) + 2\Sigma_{\text{Ch}}(0)$. The blue triangle symbols show the resulting bandgap renormalization (continuum of free e-h pairs). When the hole density is $2.5 \times 10^{12}$ cm$^{-2}$ ($N=6$), the renormalization leads to energy redshift of 230~meV compared to the one at charge neutrality. This energy renormalization is in good agreement with the experimental findings from $\mu$-scale  ARPES by Nguyen et al.  in Fig.~\ref{fig:ExpXXu} \cite{Nguyen_Nat19}. 

\section{Energy renormalization of excitons}\label{sec:BGRX}
The most common theory for the energy  renormalization of excitons assumes free e-h pairs, where the exciton energy is renormalized by the sum of self-energies of its constituent particles. Such treatment leads to significant decrease in the energy of a free e-h pair due to considerable change of its electron and hole self-energies when the density of resident carriers increases (Fig.~\ref{Fig:BothBGR}(c)). However, this is not what has been observed experimentally for tightly bound excitons (Figs. \ref{fig:ExpXXu} and \ref{fig:ExpDat2}(a)) \cite{Wang_PRL18,Back_PRL17,Forste_NatCom20,Robert_PRL21,Liu_PRL19,He_NatComm20,Lindlau_NatComm18,Liu_PRR19,Nguyen_Nat19,Liu_PRL20}. The following are several deficiencies of a  theory that treats  the energy renormalization of a bound exciton as the sum of independent electron and hole components. First, there are differences between the potential generated by an exciton and the one by a free carrier.    While the exciton is a dipole with potential rapidly decreasing  over the distance, a free electron or hole, on the other hand, is a charged particle of which Coulomb potential remains significant over a  long range, giving rise to their large energy renormalization. Second, even if we consider the exciton potential as a sum of the potentials generated by its electron and hole components, their contributions to the exciton potential are {\it destructive}. The exciton self-energy according to the theory of a free e-h pair, on the other hand, has {\it constructive} contributions from the self-energies of its components. Namely, the added contributions of the Coulomb-hole components in $\Delta E_\text{g} = \Sigma_\text{sx} + 2 \Sigma_\text{Ch}$ considerably shrink the optical gap of a free e-h pair  \cite{Haug_SchmittRink_PQE84}. Third, the theory of a free e-h pair does not consider the repeated change in components of the exciton due to the component-exchange interaction with resident charges \cite{VanTuan_PRB25}. Component exchange describes the Coulomb interaction between an exciton and resident carriers in which the exciton swaps its constituent particles  with the carriers.

Here, we argue that in order to calculate the energy renormalization of excitons, one should consider the exciton as a quasiparticle rather than a combination of two separate components. The argument is based on the fact that the ground state exciton in TMDs is a tightly bound state with binding energy in the order of hundreds of meV, much larger than the Fermi energy of the resident carriers, making it a well-defined quasiparticle \cite{Goryca_NatCom19,Chernikov_PRL14,Qiu_PRL13,VanTuan_PRB18}. By treating the exciton as a quasiparticle, we can include its component-exchange interaction with resident carriers. 
  
\begin{figure}[t] 
\centering
\includegraphics[width=8.5cm]{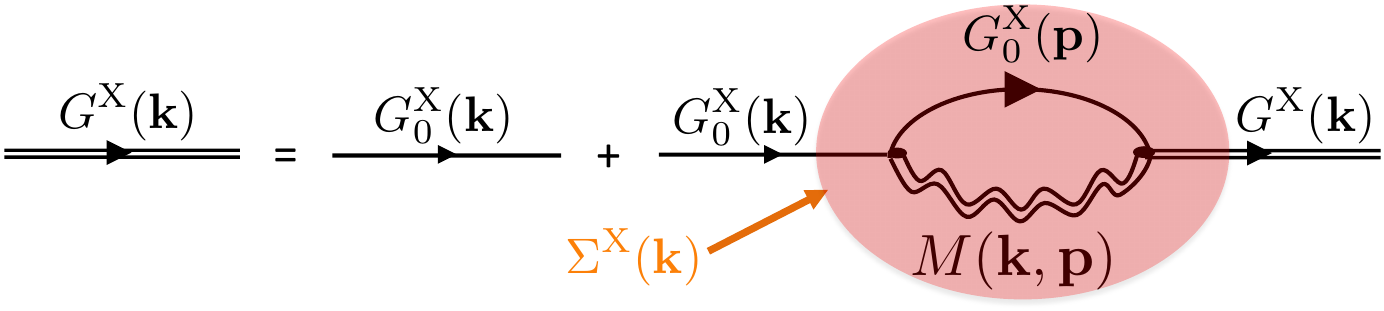}
\caption{ Feynman diagram representing the energy renormalization of an exciton due to the interaction with a Fermi sea of identical particles. The exciton self-energy $\Sigma^\text{X}({\bf k}, \omega)$ is highlighted.  }\label{fig:DysonX} 
\end{figure}
  
We consider WSe$_2$ monolayers which has resident holes with the same quantum numbers as the hole component of the exciton (Fig.~\ref{fig:ExpDat2}(c)), meaning that the exciton cannot form a trion by binding to a resident hole \cite{Dery_PRX25}. We use the GW theory to calculate the exciton energy renormalization under the effect of component exchange interaction with free holes from the Fermi sea.  The interacting Green's function of the exciton, corresponding to the  diagram shown in Fig.~\ref{fig:DysonX}, is given by
\begin{equation}
G^\text{X}({\bf k}, \omega) = \frac{1}{\hbar\omega + i \delta  - E^\text{X}_{\bf k} -\Sigma^\text{X} ({\bf k},\omega) }\,,
\end{equation}
where ${\bf k}$ is the translational momentum of the exciton, $E^\text{X}_{\bf k}$ is the energy dispersion of the exciton without resident carriers, and $\delta$ is a broadening parameter. The self-energy of the exciton reads
\begin{equation}
\Sigma^\text{X}({\bf k},\omega) = \sum_{\bf p}  M({\bf k,p}) G^\text{X}_0({\bf p},\omega) \,,
\end{equation}
 where  $G^\text{X}_0({\bf k}, \omega) = 1/(\hbar\omega + i \delta  - E^\text{X}_{\bf k})$ is the noninteracting Green's function of the exciton, and  $M({\bf k,p})$ is the matrix element representing the interaction between the exciton and resident holes. This matrix element has been derived in Ref.~\cite{VanTuan_PRB25},
\begin{eqnarray}
 M({\bf k,p}) &=& \sum_{{\bf k}_\text{h},{\bf k}'_\text{h}} \langle X^0_{\bf p},h_{{\bf k}'_\text{h}}| V_{\text{hh}}+V_{\text{eh}}| X^0_{\bf k},h_{{\bf k}_\text{h}}\rangle  \nonumber \\
 &=&  \sum_{{\bf k}_\text{h}} \left( D({\bf k,p}) + X({\bf k+k}_\text{h},{\bf k,p}) \right). 
\end{eqnarray}
$| X^0_{\bf k},h_{{\bf k}_\text{h}}\rangle $ is the state of a noninteracting exciton-hole system in which the exciton and a resident hole have momenta ${\bf k}$ and ${\bf k}_\text{h}$, respectively. We denote the Coulomb interaction between the exciton’s electron (hole) and the resident hole   by $V_{\text{eh}}$ ($V_{\text{hh}}$). The matrix element has direct ($D$) and exchange ($X$) contributions corresponding to  scattering  processes in which the exciton keeps and changes its hole component, respectively.   It has been shown that  the exchange interaction is far more dominant than the direct one  \cite{VanTuan_PRB25}. Neglecting the weak direct component, the matrix element becomes
\begin{eqnarray}
\!\!\!M({\bf k,p}) &\simeq& \sum_{{\bf k}_\text{h}} X({\bf k+k}_\text{h},{\bf k,p})   = \frac{1}{A}  \sum_{\mathbf{q},{\bf k}_\text{h}} V_{\mathbf{q}} \, \phi_{\eta\mathbf{p} -{\bf k}_\text{h} - \mathbf{q}} \nonumber \\
   &\,& \qquad \times \left[ \phi^\ast_{\eta\mathbf{k}+\mathbf{p} -{\bf k}_\text{h} }  -  \phi^\ast_{\eta\mathbf{k}+\mathbf{p} -{\bf k}_\text{h} - \mathbf{q}} \right],\,\,\, \label{Eq:MEx} 
\end{eqnarray}
where $\phi_{\bf k}$ is the exciton wavefunction in momentum space.  $\eta =m_\text{h}/m_\text{X} $ is the ratio between the hole  ($m_\text{h}$) and exciton ($m_\text{X} = m_h + m_e$) masses.  The calculations are performed by using the Rytova-Keldysh (RK) potential \cite{Rytova_MSU67,Keldysh_JETP79,Cudazzo_PRB2011}, and the exciton wavefunction is obtained from the Stochastic Variational Method in momentum space (SVM-k) \cite{VanTuan_ArXiv22,VanTuan_PRB22}.  Appendix \ref{app:ExNorm} gives detailed calculations for the matrix element  $M({\bf k,p})$ using the SVM-k wavefunction.  

\begin{figure}[t] 
\centering
\includegraphics[width=8.5cm]{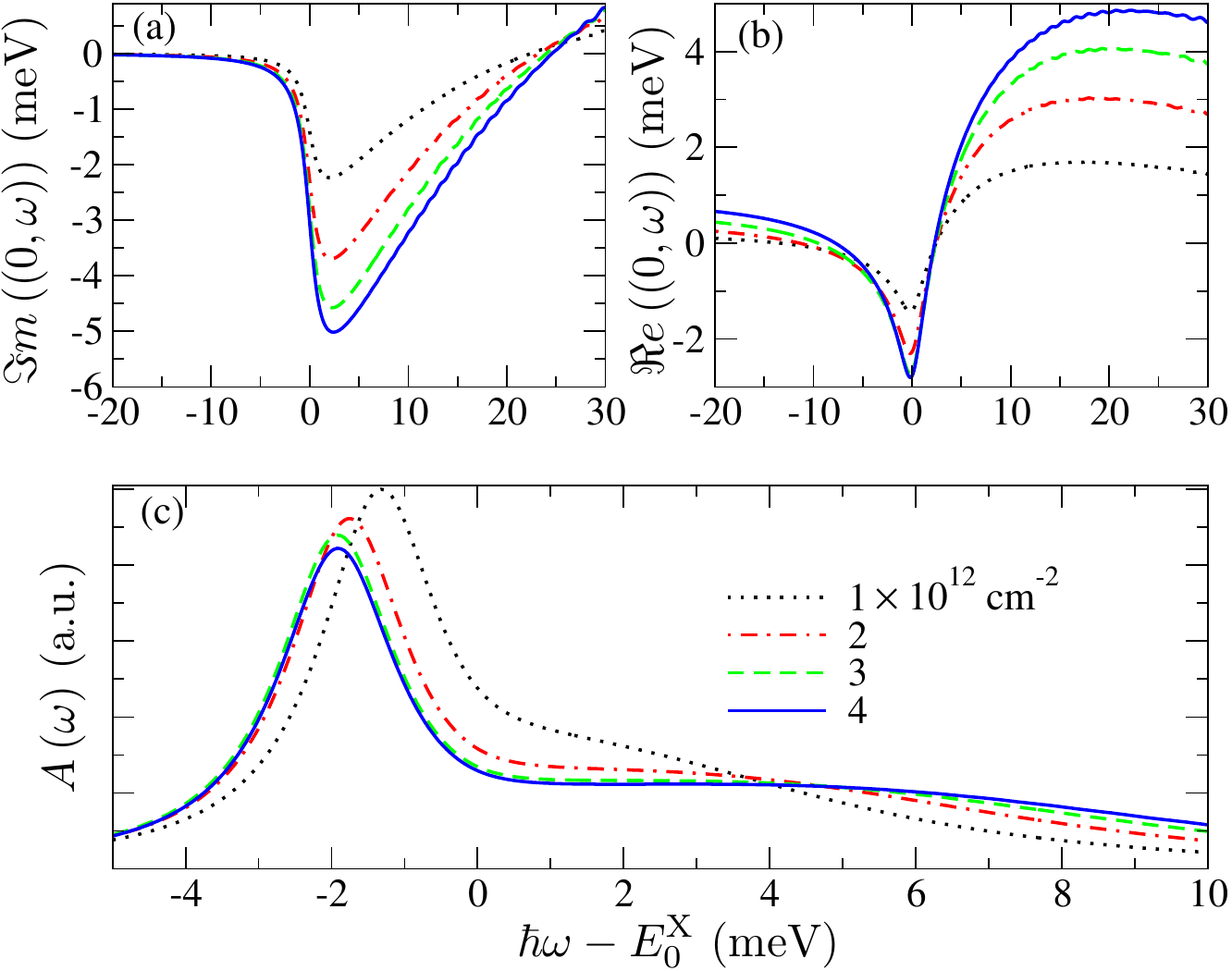}
\caption{ The imaginary (a), real (b) parts of the exciton self-energy, and the spectral function  (c) as a function of $\hbar \omega - E^\text{X}_0$. The three quantities are determined at zero exciton momentum, ${\bf k}=0$. The parameters used in the calculations are detailed in Appendix \ref{app:num}.   }\label{fig:ExcitonR} 
\end{figure}

Figures~\ref{fig:ExcitonR}(a) and (b) show the imaginary and real parts, respectively, of the self-energy of  an exciton in the light cone, $\Sigma^\text{X}({\bf k}=0,\omega)$. The small energy renormalization of the exciton can be realized from the few meV shift of the resonance energy when the resident hole density is a few $10^{12}$ cm$^{-2}$. The zero energy of the $x$-axis is the exciton resonance energy without resident holes. To study the energy renormalization through the spectral position of the exciton, we calculate the spectral function
\begin{equation}
A(\omega) \propto -  \Im m\left(  G^\text{X}({\bf k} = 0, \omega) \right)\, ,
\end{equation} 
and show the results in Fig.~\ref{fig:ExcitonR}(c). The exciton resonance redshifts less than $2$~meV when the hole density is $4 \times 10^{12}$ cm$^{-2}$. Such small redshift is overshadowed by  the band filling effects, explaining the  blueshifts observed in Fig.~\ref{fig:ExpDat2}(a).  The smallness of the exciton energy renormalization is a consequence of the dipole nature of excitons manifested  by the destructive contributions of  $V_{\text{eh}}$ and $V_{\text{hh}}$, denoted by the terms in square brackets of Eq.~(\ref{Eq:MEx}). Physically, the insensitivity of the exciton resonance to the density of resident carriers (in the studied   range) is explained by the small exciton radius ($\sim 1$~nm) and the short-range nature of the potential generated by its dipole, which decreases considerably at a distance of few nm between resident carriers.

\section{Summary}\label{sec:sum}

We have analyzed the contributions to energy renormalization of resident carriers in magnetic field  and compared the theoretical results  with experimental observations.  The
 findings highlighted the important role of dynamical screening through the Coulomb-hole energy in TMD monolayers. It has been shown that the RPA, although slightly overestimating the screening,   still provides a good description for the energy renormalization of resident carriers in TMD monolayers. Comparing the energy renormalization of resident carriers with the energy shift of excitonic resonances, we have seen that the measured shift cannot be explained by adding the energy renormalization of independent electron and hole  components.  A theoretical model in which the exciton is treated as a quasiparticle has been developed to study its energy renormalization due to interaction with resident carriers. The obtained results explain the weak energy shift of the excitonic resonances when resident carriers are added to the monolayer.
 
\acknowledgments{ This work is supported by the Department of Energy, Basic Energy Sciences, Division of Materials Sciences and Engineering under Award No. DE-SC0014349.}
\appendix
\section{Screened exchange energy  } \label{app:SX}

\begin{figure}[h] 
\centering
\includegraphics[width=8.5cm]{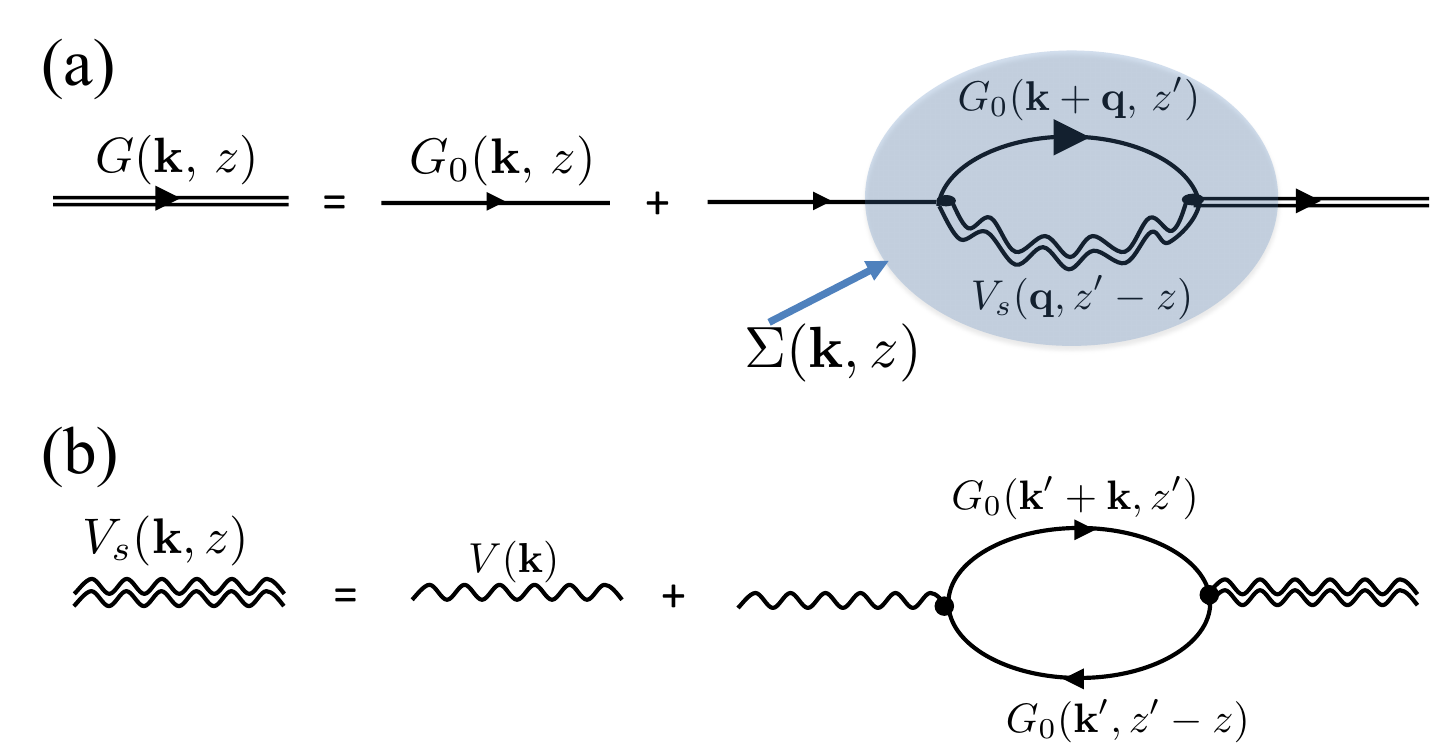}
\caption{ Feynman diagram representing Dyson equation for (a) electron Green's function  and  (b) screened Coulomb potential. The structure of the electron self-energy is highlighted in the shaded oval in (a).   }\label{fig:Dyson} 
\end{figure}

Figure \ref{fig:Dyson}(a) shows the Feynman diagram for the interacting electron Green's function $G({\bf k}, z)$, satisfying the Dyson equation
\begin{equation}
\!\!\!\!\!\!G({\bf k},\!z) \!\!= \!\! \frac{G_0({\bf k}, \!z)}{1 - G_0({\bf k},\!z) \Sigma ({\bf k},\!z)}\!\! =\!\! \frac{1}{z - \varepsilon_{\bf k}+ \mu -\Sigma ({\bf k},\!z) }\,,
\end{equation}
where $G_0({\bf k}, z) = \left( z - \varepsilon_{\bf k}+ \mu\right)^{-1}$ is the non-interacting Green's function, and $\mu$ is the electron chemical potential. $z$ is the fermionic Matsubara frequency.  The self-energy is given by
\begin{equation}
\Sigma ({\bf k},z) = -\frac{1}{\beta}\sum_{{\bf q},z'} G_0({\bf k+q}, z') V_s({\bf q}, z' - z)\,,
\label{Eq:SelfEDyna}
\end{equation}
where $\beta = \left(k_\text{B} T \right)^{-1}$ and $T$ is the temperature of the electron system. $V_s({\bf q}, z' - z)$ is the dynamically screened potential, which in the RPA framework, described by the diagram in Fig.~\ref{fig:Dyson}(b)
\begin{equation}
V_s({\bf q},\omega) = \frac{V({\bf q})}{1-V({\bf q}) \chi_0({q},\omega)} = \frac{V({\bf q})}{\epsilon_s({q},\omega)}\,.
\end{equation}
$V({\bf q})$ is the bare Coulomb potential, and the screened dielectric function reads 
\begin{equation}
\epsilon_s({q},\omega) = 1-V({\bf q}) \chi_0({q},\omega)\,,
\end{equation}
where $\chi_0({q},\omega)$ is the non-interacting response function \cite{Giuliani_Vignale_Book}. 

If we neglect the dynamical effects by using the static limit of the screened potential, i.e., $V_s({\bf q}, z' - z)\simeq V_s({\bf q}, 0)$, the self-energy in Eq.(\ref{Eq:SelfEDyna}) becomes the screened-exchange energy
\begin{equation}
\Sigma_\text{sx} ({\bf k}) =  -  \sum_{{\bf q}} f({\bf k+q}) V_s({\bf q})\,,
\label{Eq:sxE}
\end{equation}
where $V_s({\bf q})  = V_s({\bf q},\omega = 0) $  is the static screened potential and  $f({\bf k})$ is the Fermi-Dirac distribution.

\section{Coulomb-hole  energy } \label{app:CH}
The dynamical effects has been considered  in several references. It is important for accurately describing  many-body effects as well as optical properties of TMD monolayers \cite{VanTuanPRB24,Mhenni_ACS25,VanTuan_PRX17}. Here, we substitute the spectral representation of the screened Coulomb potential \cite{Haug_SchmittRink_PQE84,HaugBook}
\begin{equation}
V_s({\bf q}, z' - z) = V({\bf q}) - \int_{-\infty}^{\infty} \frac{d \omega}{\pi} \frac{\Im m V_s({\bf q},\omega) }{z' - z - \hbar\omega},
\end{equation}
into Eq.~\ref{Eq:SelfEDyna}, and then perform the sum over the Matsubara frequencies $z'$. Next, we perform analytic continuation $z \rightarrow \varepsilon_{\bf k} -\mu + i\delta $ to the real-energy axis. By neglecting the  recoil energy (see Refs.~\cite{Haug_SchmittRink_PQE84,Scharf_JPCM19} for more details), we finally obtain 
 \begin{equation}
\Sigma ({\bf k})  = \Sigma_\text{sx} ({\bf k})   + \Sigma_\text{Ch} ({\bf k}). 
 \end{equation}
The Coulomb-hole component of the self-energy is the additional contribution from the dynamical effects
 \begin{equation}
 \Sigma_\text{Ch} ({\bf k})  = - \frac{1}{2} \sum_{\bf q} \left(V({\bf q}) - V_s({\bf q}) \right). 
 \end{equation}

\section{Screened-exchange and Coulomb-hole energies in a 2D electron gas under magnetic field} \label{app:SelfEB}
We generalize the  theory to the case of 2D electron gas  in magnetic field $\bf B$. The fact that the formalism above does not specify the energy dispersion, $\varepsilon_{\bf k}$, means that the theory is also applicable  in the case of electrons in magnetic field. In the Landau-gauge basis, the quantum number  becomes $ {\bf k} \rightarrow \{ n, k\}$ 
 where $n$ is the Landau level index and $k$ is the electron wave vector along the $y$ direction. The Coulomb potential  becomes a matrix with elements \cite{VanTuanPRB25_2}
  \begin{eqnarray} 
V_{n_1,k_1;n_2,k_2 }^{n'_1,k'_1;n'_2,k'_2 } &=& \langle n'_1,k'_1; n'_2,k'_2| \hat{V}| n_1,k_1; n_2,k_2 \rangle \nonumber \\
&=&  \!\sum_{\bf q} e^{i( k_1 - k_2)q_x \ell_\text{B}^2} \,\, W_{n_1;n_2 }^{n'_1;n'_2 }({\bf q}) \delta_{k_1;k_2}^{k'_1;k'_2}\!(q_y) \,,\,\,\,\, \,\,\,\,\,\,\,\,\,\,
\end{eqnarray}
where ${\bf q} = (q_x,q_y)$ and $\ell_\text{B} = \sqrt{\hbar/(eB)}$ are the transferred momentum and magnetic length, respectively.  $ \delta_{k_1;k_2}^{k'_1;k'_2}(q_y) \equiv \delta_{k'_1,k_1+q_y} \delta_{k'_2,k_2-q_y} $ stems from momentum conservation of the two interacting electrons. The matrix element $W_{n_1;n_2 }^{n'_1;n'_2 }({\bf q})$ is given by
\begin{equation}
W_{n_1;n_2 }^{n'_1;n'_2 }({\bf q})  = V({\bf q}) \,\, S_{n_1}^{n'_1}({\bf q}) \,\, S_{n_2}^{n'_2}({- \bf q}), 
\label{Eq:Wq}
\end{equation}
where the form factor function reads
\begin{equation}
S_{n}^{n'}({\bf q})  = i^{|\Delta n|}\,\, e^{  \frac{i}{2} q_x q_y \ell^2_\text{B}+ i \Delta n \theta} \,\, \tilde{L}^{|\Delta n|}_{\tilde{n}} \left( \frac{q^2 \ell_\text{B}^2}{2} \right).
\label{Eq:Sq}
\end{equation}
$\Delta n  = n' - n$ and $ \tilde{n}= \text{min}\!\left\{n,n' \right\}$  are the difference and minimum of the two LL indices, respectively. $\theta =\arctan\left( q_y/q_x\right)  $ is the angle between ${\bf q}$ and the $x$-axis, and the functions $ \tilde{L}^m_n(x)$  were defined in Eq.~(\ref{Eq:Lmntilde}). 

The screened exchange energy in Eq.~(\ref{Eq:sxE}) becomes 
\begin{eqnarray}
\Sigma_\text{sx}(n) &=& -\sum_{n',k'} f_{n'} V_{n, k;n',k'}^{n',k'; n,k }    \nonumber \\
&=& -\sum_{n',k'}  f_{n'}   \sum_{\bf q} e^{i( k - k')q_x \ell_\text{B}^2} \,\, W_{n;n' }^{n';n}({\bf q}) \delta_{k;k'}^{k';k}\!(q_y) \nonumber \\
&=& -\sum_{n'}  f_{n'}   \sum_{\bf q} W_{n;n' }^{n';n}({\bf q})  \sum_{k'} e^{i( k - k')q_x \ell_\text{B}^2} \,\,  \delta_{k;k'}^{k';k}\!(q_y)\nonumber \\
&=& -\sum_{n'}  f_{n'}   \sum_{\bf q} W_{n;n' }^{n';n}({\bf q})   \,\,\, e^{-iq_x q_y \ell_\text{B}^2} \nonumber \\
&=&  -\sum_{n'}  f_{n'}   \sum_{\bf q} V_s({\bf q}) \,\,\, \left( \tilde{L}^{|\Delta n|}_{\tilde{n}} \left( \frac{q^2 \ell_\text{B}^2}{2} \right) \right)^2 ,
\end{eqnarray}
which is Eq.~(\ref{Eq:SelfSX}) of the main text.

Similarly, we can obtain the Coulomb-hole energy in Eq.~(\ref{Eq:SelfCh})  for an electron in the LL $n$  as
\begin{equation}
\Sigma_\text{Ch}(n)
=  -\frac{1}{2}    \sum_{n',{\bf q}} \left[ V({\bf q}) \!\!- \!\!V_s({\bf q}) \right] \left( \tilde{L}^{|\Delta n|}_{\tilde{n}} \!\!\left( \frac{q^2 \ell_\text{B}^2}{2} \right) \right)^2 .
\end{equation}

\section{Detailed self-energy calculations of excitons using SVM-k} \label{app:ExNorm}
The  wave function for the relative motion between the electron and hole of an exciton is obtained from the SVM-k model \cite{VanTuan_ArXiv22}
\begin{equation}
  \phi^\text{SVM}_{\bf k} = \sum^{\mathcal{K}}_i C_i e^{-\frac{1}{2}M_ik^2}  \,, \label{Eq:WaveSVM}
\end{equation}
where $\mathcal{K}$ is the number of basis functions used to construct the exciton wavefunction.   $M_i$ are variational parameters, which together with the coefficients $C_i$, obtained from minimization of the exciton energy \cite{VanTuan_ArXiv22,VanTuan_PRB22}.  The results in Fig.~\ref{fig:ExcitonR} are obtained from calculations of the wave function in Eq.~(\ref{Eq:WaveSVM}), and the Rytova-Keldysh potential 
\begin{equation}
V({\bf q}) = \frac{2 \pi e^2}{  Aq \epsilon_s(1+r_0q)}.
\label{Eq:VRK}
\end{equation}
$A$ is sample area, $\epsilon_s$ is the dielectric constant of the environment surrounding the TMD monolayer, and $r_0$ is  the polarizability of the TMD monolayer \cite{Cudazzo_PRB2011}. 
 The potential parameters used in the calculations are listed in Appendix \ref{app:num}. The matrix element in Eq.(\ref{Eq:MEx}) can be rewritten as
\begin{eqnarray}
M({\bf k,p}) =   \sum_{{\bf k}_\text{h}}  F({\bf k, p,k}_\text{h})\,,
\end{eqnarray}
where
\begin{eqnarray}
 F({\bf k, p,k}_\text{h}) &=&     \sum_{{\bf q}}      V_{\bf q}  \,\,\, \phi_{{\bf Q}_1 - \mathbf{q}}   \left[ \phi^\ast_{\mathbf{Q}_2 }  -  \phi^\ast_{{\bf Q}_2 - \mathbf{q}} \right], 
\end{eqnarray}
with ${\bf Q}_1 = \eta\mathbf{p} -{\bf k}_\text{h}$ and $ {\bf Q}_2 = \eta\mathbf{k}+\mathbf{p} -{\bf k}_\text{h}$. 
\begin{widetext}
 \begin{eqnarray}
 &\,& F({\bf k, p,k}_\text{h}) =  \nonumber \\ &\,& -\sum_{i,j} C_i C_j  \frac{A}{4 \pi^2}  \int d{\bf q}  \frac{2\pi e^2}{A q \epsilon_0 (1+ r_0 q)} \exp \left(- \frac{M_i}{2}\left( {\bf Q}_1 - {\bf q}\right)^2 \right) \left[ \exp \left( -\frac{M_j}{2}\left( {\bf Q}_2 - {\bf q}\right)^2 \right) -  \exp \left( -\frac{M_j {\bf Q}_2^2 }{2} \right) \right] \,\,\, \nonumber \\
&=& - \sum_{i,j}   \frac{e^2 C_i C_j}{2 \pi  \epsilon_0}  \exp \left(-\frac{M_i {\bf Q}_1^2 }{2} -\frac{M_j {\bf Q}_2^2 }{2} \right)  \int   \frac{ d{\bf q} }{ q  (1+ r_0 q)} \left[ \exp \left( -M_{ij} {\bf q}^2 + {\bf P}_{ij} {\bf q}\right) -  \exp \left(- \frac{M_i}{2} {\bf q}^2 + M_i{\bf Q}_1 {\bf q} \right)  \right] \,\,\, \nonumber \\
&=& - \sum_{i,j}   \frac{e^2 C_i C_j}{  \epsilon_0 r_0}  \exp \left(-\frac{M_i {\bf Q}_1^2 }{2} -\frac{M_j {\bf Q}_2^2 }{2} \right)  \int_0^{\infty}   \frac{ dx }{   1+ x} \left[ \exp \left( -\tilde{M}_{ij} x^2 \right)I_0 \left(\tilde{P}_{ij} x\right) -  \exp \left(- \frac{\tilde{M}_i}{2} x^2\right) I_0 \left( M_i\tilde{Q}_1  x\right)  \right] . \,\,\, \nonumber \\
 \label{Eq:MatEleKeldysh}
 \end{eqnarray} 
 \end{widetext}
$I_0(x)$ is the modified Bessel functions of the  first  kind, and other variables are $ M_{ij} = (M_i+M_j)/2$,  ${\bf P}_{ij} = M_i{\bf Q}_1 + M_j{\bf Q}_2$, $ \tilde{M}_{ij} =M_{ij}/r_0^2 $, $\tilde{P}_{ij} = P_{ij} /r_0$,    $ \tilde{M}_{i} =M_{i}/r_0^2 $, and $\tilde{Q}_1 = Q_1 / r_0$.

\section{Parameters} \label{app:num}
The calculations in this work use the following parameters for hBN-encapsulated WSe$_2$ monolayer.  
\begin{enumerate}
\item In the case of energy renormalization of electron-doped WSe$_2$ monolayer in the bottom CB valley,  we use the effective mass  $m_\text{c} = 0.4 m_0$  \cite{Kormanyos_2DMater15}, and  $g$-factor $g_\text{c} = 0.86$ \cite{Robert_PRL21}.

\item In the case of energy renormalization of  hole-doped WSe$_2$ monolayer in the top VB valley,  we use the effective mass $m_\text{v} = 0.36 m_0$  \cite{Kormanyos_2DMater15}  and   $g$ factor $g_\text{v} = 6.1$ \cite{Robert_PRL21}.

\item  The Rytova-Keldysh potential  is described by  Eq.~(\ref{Eq:VRK}) in which $\epsilon_s = 3.8$ is the dielectric constant of hBN  \cite{VanTuan_PRB18,Cai_SSC2007,Dai_Science2014}, and $r_0=r_0^*/\epsilon$ with  the polarizability of the TMD monolayer $ r_0^* = 4.5$ nm \cite{Stier_PRL18}.

\item In the case of energy renormalization of excitons in hole-doped WSe$_2$ monolayer,  we use the effective masses  $m_\text{h} = 0.36 m_0$,  $m_\text{e} = 0.29 m_0$  \cite{Kormanyos_2DMater15} in the VB and top CB, respectively.   The exciton energy dispersion is $E^\text{X}_{\bf k} = \hbar^2 k^2/2m_\text{X}$, where the exciton  mass is given by $m_\text{X}   = m_\text{e} + m_\text{h}$. The broadening parameter in the Green's function is $\delta = 1$~meV.

\end{enumerate}


\begin{thebibliography}{99}

\bibitem{Wang_RMP18}  G. Wang, A. Chernikov, M. M. Glazov, T. F. Heinz, X. Marie, T. Amand, and B. Urbaszek, Colloquium: Excitons in atomically thin transition metal dichalcogenides, Rev. Mod. Phys. \textbf{90}, 021001 (2018).

\bibitem{Cudazzo_PRB11} P. Cudazzo, I. V. Tokatly, and A. Rubio, Dielectric screening in two-dimensional insulators: Implications for excitonic and impurity states in graphene, Phys. Rev. B {\bf 84}, 085406 (2011).

\bibitem{Meckbach_PRB18} L. Meckbach, T. Stroucken, and S. W. Koch, Influence of the effective layer thickness on the ground-state and excitonic properties of transition-metal dichalcogenide systems, Phys. Rev. B \textbf{97}, 035425 (2018).

\bibitem{VanTuan_PRB18} D. V. Tuan, M. Yang, and H. Dery, Coulomb interaction in monolayer transition-metal dichalcogenides, Phys. Rev. B \textbf{98}, 125308 (2018).

\bibitem{Marauhn_PRB23} P. Marauhn and M. Rohlfing, Image charge effect in layered materials: Implications for the interlayer coupling in MoS$_2$, Phys. Rev. B \textbf{107}, 155407 (2023).

\bibitem{Nuckolls_NatRevMat24}
K. P. Nuckolls and  A. Yazdani, A microscopic perspective on moiré materials, Nat. Rev. Mater. {\bf 9},  460-480 (2024).

\bibitem{Andrei_NatRevMat21}
E. Y. Andrei, D. K. Efetov, P. Jarillo-Herrero, A. H. MacDonald, K. F. Mak, T. Senthil, E. Tutuc, A. Yazdani, and A. F. Young, The marvels of moiré materials, Nat. Rev. Mater. {\bf  6}, 201-206 (2021). 

\bibitem{Du_Science23}
L. Du, M. R. Molas, Z. Huang, G. Zhang, F. Wang, Z. Sun, Moir\'{e} photonics and optoelectronics, Science {\bf 379},  6639 (2023). 

\bibitem{Paik_AOP24}
E. Paik, L. Zhang, K. F. Mak, J. Shan, and H. Deng, Excitons and polaritons in two-dimensional transition metal dichalcogenides: a tutorial, Advances in Optics and Photonics, {\bf 16}(4), 1064-1132 (2024).

\bibitem{Ju_NatRevMater24}
L. Ju, A. H. MacDonald, K. F. Mak, J. Shan, and  X. Xu, The fractional quantum anomalous Hall effect, Nat. Rev. Mater.  {\bf 9}, 455-459 (2024). 

\bibitem{Abajo_ACS25}
F. Javier Garcia de Abajo \textit{et al.}, Roadmap for photonics with 2D materials,  ACS Photonics, {\bf 12}, 3961-4095 (2025).

\bibitem{Bernevig_NatPhys25}
B. A. Bernevig, L. Fu, L. Ju, A. H. MacDonald, K. F. Mak, and J. Shan, Fractional quantization in insulators from Hall to Chern, Nat. Phys. {\bf 21}, 1702-1713 (2025).   

\bibitem{Xia_Nat26}
Y. Xia, Z. Han, J. Zhu, Y. Zhang, P. Knuppel, K. Watanabe, T. Taniguchi, K. F. Mak, and J. Shan,  Bandwidth-tuned Mott transition and superconductivity in moir\'{e} WSe$_2$. Nature  {\bf 650}, 585-591 (2026).

\bibitem{Xia_Nat25}
Y. Xia, Z. Han, K. Watanabe, T. Taniguchi, J. Shan, and K. F. Mak,  Superconductivity in twisted bilayer WSe$_2$. Nature {\bf 637},  833-838 (2025).

\bibitem{Liu_PRL20} E. Liu, J. van Baren, T. Taniguchi, K. Watanabe, Y.-C. Chang, and C. H. Lui, Landau-quantized excitonic absorption and luminescence in a monolayer valley semiconductor, Phys. Rev. Lett. \textbf{124}, 097401 (2020).


\bibitem{Wang_NanoLett17} Z. Wang, L. Zhao, K. F. Mak, and J. Shan, Probing the spin-polarized electronic band structure in monolayer transition metal dichalcogenides by optical spectroscopy, Nano Lett. \textbf{17}, 740 (2017).

\bibitem{VanTuan_PRL22}  D. V. Tuan, S.-F. Shi, X. Xu, S. A. Crooker, and H. Dery,  Six-body and eight-body exciton states in monolayer WSe$_2$, Phys. Rev. Lett. \textbf{129}, 076801 (2022).

\bibitem{Dijkstra_NatCom25} A. Dijkstra, A. B. Mhenni, D. V. Tuan, E. Cetiner, M. Schur-Wilkens, J. Kim, L. Steiner, K. Watanabe, T. Taniguchi, M. Barbone, N. P. Wilson, H. Dery, J. J. Finley, Ten-valley excitonic complexes in charge-tunable monolayer WSe$_2$, Nat. Commun. {\bf 16}, 9743 (2025).

\bibitem{Jindal_PRB25}
V. Jindal, K. Mourzidis, A. Balocchi, C. Robert, P. Li, D. Van Tuan, L. Lombez, D. Lagarde, P. Renucci, T. Taniguchi, K. Watanabe, H. Dery, and X. Marie, Brightened emission of dark trions in transition metal dichalcogenide monolayers, Phys. Rev. B {\bf 111}, 155409 (2025).  

\bibitem{Ren_PRB23} L. Ren, C. Robert, H. Dery, M. He, P. Li, D. V. Tuan, P. Renucci, D. Lagarde, T. Taniguchi, K. Watanabe, X. Xu, and X. Marie, Measurement of the conduction band spin-orbit splitting in  WSe$_2$ and  WS$_2$ monolayers, Phys. Rev. B  \textbf{107}, 245407 (2023).

\bibitem{Kapuncinski_CP21} P. Kapu\'{s}ci\'{n}ski, A. Delhomme, D. Vaclavkova, A. O. Slobodeniuk, M. Grzeszczyk, M. Bartos, K. Watanabe, T. Taniguchi, C. Faugeras, and M. Potemski, Rydberg series of dark excitons and the conduction band spin-orbit splitting in monolayer WSe2, Commun. Phys. \textbf{4}, 186 (2021).


\bibitem{Mahan_Book} G. D. Mahan, \textit{Many-particle physics}, 2$^\text{nd}$ Edition, Plenum Press, New York (1990).

\bibitem{Giuliani_Vignale_Book} G. Giuliani and G. Vignale, \textit{Quantum theory of the electron liquid} (Cambridge University Press, Cambridge, 2005).

\bibitem{Krishtopenko_JPCM11} S. S. Krishtopenko, V. I. Gavrilenko and M. Goiran, Theory of $g$-factor enhancement in narrow-gap quantum well heterostructures, J. Phys. Condens. Matter \textbf{23}, 385601 (2011).

\bibitem{Oreszczuk_2DMater13} K. Oreszczuk, A. Rodek, M. Goryca, T. Kazimierczuk, M. Raczy\'{n}ski, J. Howarth, T. Taniguchi, K Watanabe, M. Potemski, and P. Kossacki, Enhancement of electron magnetic susceptibility due to many-body interactions in monolayer MoSe$_2$, 2D Mater. \textbf{10}, 045019 (2023).



\bibitem{Back_PRL17}
P. Back, M. Sidler, O. Cotlet, A. Srivastava, N. Takemura, M. Kroner, and A. Imamoglu, Giant paramagnetism-induced valley polarization of electrons in charge-tunable monolayer MoSe$_2$, Phys. Rev. Lett. {\bf 118}, 237404 (2017). 

\bibitem{Xu_PRL17}
S. Xu, J. Shen, G. Long, Z. Wu, Z. Bao, C.-C. Liu, X. Xiao, T. Han, J. Lin, Y. Wu, H. Lu, J. Hou, L. An, Y. Wang, Y. Cai, K. M. Ho, Y. He, R. Lortz, F. Zhang, and N. Wang, Odd-integer quantum Hall states and giant spin susceptibility in $p$-type few-layer WSe$_2$, Phys. Rev. Lett. {\bf 118}, 067702 (2017). 

\bibitem{Nedniyom_PRB09} B. Nedniyom, R. J. Nicholas, M. T. Emeny, L. Buckle, A. M. Gilbertson, P. D. Buckle, and T. Ashley, Giant enhanced g-factors in an InSb two-dimensional gas, Phys. Rev. B \textbf{80}, 125328 (2009).

\bibitem{Larentis_PRB18} S. Larentis, H. C. P. Movva, B. Fallahazad, K. Kim, A. Behroozi, T. Taniguchi, K. Watanabe, S. K. Banerjee, and E. Tutuc, Large effective mass and interaction-enhanced Zeeman splitting of K-valley electrons in MoSe$_2$, Phys. Rev. B \textbf{97}, 201407(R)  (2018).

\bibitem{Wang_PRL18}
Z. Wang, K. F. Mak, and J. Shan, Strongly interaction-enhanced valley magnetic response in monolayer WSe$_2$, Phys. Rev. Lett. {\bf 120}, 066402 (2018). 

\bibitem{Mak_PRL10} K. F. Mak, C. Lee, J. Hone, J. Shan, and T. F. Heinz, Atomically thin MoS$_2$: A new direct-gap semiconductor, Phys. Rev. Lett. \textbf{105}, 136805 (2010).

\bibitem{Splendiani_NanoLett10} A. Splendiani, L. Sun, Y. Zhang, T. Li, J. Kim, C.-Y. Chim, G. Galli, and F. Wang, Emerging photoluminescence in monolayer MoS$_2$, Nano Lett. \textbf{10}, 1271 (2010).

\bibitem{Chernikov_PRL14} A. Chernikov, T. C. Berkelbach, H. M. Hill, A. Rigosi, Y. Li, O. B. Aslan, D. R. Reichman, M. S. Hybertsen, and T. F. Heinz, Exciton binding energy and nonhydrogenic Rydberg series in monolayer WS$_2$, Phys. Rev. Lett. \textbf{113}, 076802 (2014).

\bibitem{He_PRL14} K. He, N. Kumar, L. Zhao, Z. Wang, K. F. Mak, H. Zhao, and J. Shan, Tightly bound excitons in monolayer WSe$_2$, Phys. Rev. Lett. \textbf{113}, 026803 (2014).

\bibitem{Goryca_NatCom19}
M. Goryca, J. Li, A. V. Stier, T. Taniguchi, K. Watanabe, E. Courtade, S. Shree, C. Robert, B. Urbaszek, X. Marie, and S. A. Crooker, Revealing exciton masses and dielectric properties of monolayer semiconductors with high magnetic fields, Nat. Commun. {\bf 10}, 4172 (2019).

\bibitem{Mak_NatMater13} K. F. Mak, K. He, C. Lee, G. H. Lee, J. Hone, T. F. Heinz, and J. Shan, Tightly bound trions in monolayer MoS$_2$, Nat. Mater.  \textbf{12}, 207 (2013).

\bibitem{Scharf_JPCM19} B. Scharf, D. Van Tuan,  I. {\u Z}uti{\'c}, and H. Dery, Dynamical screening of excitons in monolayer transition-metal dichalcogenides, J. Phys. Condens. Matter. \textbf{31}, 203001 (2019).

\bibitem{Qiu_PRL13}
D. Y. Qiu, F. H. da Jornada, and S. G. Louie, Optical spectrum of MoS$_2$: Many-body effects and diversity of exciton states, Phys. Rev. Lett. {\bf 111}, 216805
(2013).

\bibitem{Chen_NatComm18} S.-Y. Chen, T. Goldstein, T. Taniguchi, K. Watanabe, and J. Yan,  Coulomb-bound four- and five-particle intervalley states in an atomically-thin semiconductor, Nat. Commun. {\bf 9}, 3717 (2018).

\bibitem{Ye_NatComm18} Z. Ye, L. Waldecker, E. Y. Ma, D. Rhodes, A. Antony, B. Kim, X.-X. Zhang, M. Deng, Y. Jiang, Z. Lu, D. Smirnov, K. Watanabe, T. Taniguchi, J. Hone, and  T. F. Heinz, Efficient generation of neutral and charged biexcitons in encapsulated WSe$_2$ monolayers,  Nat. Commun. {\bf 9}, 3718 (2018).

\bibitem{Li_NatComm18} Z. Li, T. Wang, Z. Lu, C. Jin, Y. Chen, Y. Meng, Z. Lian, T. Taniguchi, K. Watanabe, S. Zhang, D. Smirnov, and S.-F. Shi,  Revealing the biexciton and trion-exciton complexes in BN encapsulated WSe$_2$, Nat. Commun. {\bf 9}, 3719 (2018).  

\bibitem{Barbone_NatComm18} M. Barbone, A. R.-P. Montblanch, D. M. Kara, C. Palacios-Berraquero, A. R. Cadore, D. De Fazio, B. Pingault, E. Mostaani, H. Li, B. Chen, K. Watanabe, T. Taniguchi, S. Tongay, G. Wang, A. C. Ferrari, and M. Atat\"ure ,  Charge-tuneable biexciton complexes in monolayer WSe$_2$, Nat. Commun. {\bf 9}, 3721 (2018).

\bibitem{He_NatComm20} M. He, P. Rivera, D. V. Tuan, N. P. Wilson, M. Yang, T. Taniguchi, K. Watanabe, J. Yan, D. G. Mandrus, H. Yu, H. Dery, W. Yao, and X. Xu, Valley phonons and exciton complexes in a monolayer semiconductor, Nat. Commun. \textbf{11}, 618 (2020).

\bibitem{Liu_PRL20b} E. Liu, J. van Baren, C.-T. Liang, T. Taniguchi, K. Watanabe, N. M. Gabor, Y.-C. Chang, and C.-H. Lui, Multipath optical recombination of intervalley dark excitons and trions in monolayer WSe$_2$, Phys. Rev. Lett. \textbf{124}, 196802 (2020).

\bibitem{Choi_PRB24} J. Choi, J. Li, D. Van Tuan, H. Dery, and S. A. Crooker, Emergence of composite many-body exciton states in WS$_2$ and MoSe$_2$ monolayers, Phys. Rev. B {\bf 109}, L041304 (2024)


\bibitem{Dery_PRX25} H. Dery, C. Robert, S. A. Crooker, X. Marie, and D. V. Tuan, Energy shifts and broadening of excitonic resonances in electrostatically doped semiconductors, Phys. Rev. X \textbf{15}, 031049 (2025).



 \bibitem{Nguyen_Nat19} P. V. Nguyen, N. C. Teutsch, N. P. Wilson, J. Kahn, X. Xia, A. J. Graham, V. Kandyba, A. Giampietri, A. Barinov, G. C. Constantinescu, N. Yeung, N. D. M. Hine, X. Xu, D. H. Cobden, and  N. R. Wilson, Visualizing electrostatic gating effects in two-dimensional heterostructures,  Nature {\bf 572},  220-223 (2019). 


\bibitem{Robert_PRL21} C. Robert, H. Dery, L. Ren, D. Van Tuan, E. Courtade, M. Yang, B. Urbaszek, D. Lagarde, K. Watanabe, T. Taniguchi, T. Amand, and X. Marie, Measurement of conduction and valence bands $g$-factors in a transition metal dichalcogenide monolayer, Phys. Rev. Lett. \textbf{126}, 067403 (2021).
\bibitem{Li_PRL20}
J. Li, M. Goryca, N. P. Wilson, A. V. Stier, X. Xu, and S. A. Crooker, Spontaneous valley polarization of interacting carriers in a monolayer semiconductor, Phys. Rev. Lett. {\bf 125}, 147602 (2020). 


\bibitem{Courtade_PRB17} E. Courtade, M. Semina, M. Manca, M. M. Glazov, C. Robert, F. Cadiz, G. Wang, T. Taniguchi, K. Watanabe, M. Pierre, W. Escoffier, E. L. Ivchenko, P. Renucci, X. Marie, T. Amand, and B. Urbaszek, Charged excitons in monolayer WSe$_2$: experiment and theory, Phys. Rev. B \textbf{96}, 085302 (2017).



\bibitem{Liu_PRL19}
E. Liu, J. v. Baren, Z. Lu, M. M. Altaiary, T. Taniguchi, K. Watanabe, D. Smirnov, and C. H. Lui, Gate tunable dark trions in monolayer WSe$_2$,   Phys. Rev. Lett. \textbf{123}, 027401 (2019). 

\bibitem{Peng_RPB25}
L. Peng, G. Vignale, and S. Adam, Many-body perturbation theory for moir\'{e} systems, Phys. Rev. B {\bf 112}, 075146 (2025)



 \bibitem{VanTuanPRB25_2}
 D. V. Tuan and H. Dery, Landau-level composition of bound exciton states in magnetic field,  Phys. Rev. B {\bf 112}, 085305 (2025). 



 \bibitem{VanTuanPRB24}
 D. V. Tuan and H. Dery, Effects of dynamical dielectric screening on the excitonic spectrum of monolayer semiconductors,  Phys. Rev. B {\bf 110}, 125301 (2024). 

\bibitem{Mhenni_ACS25} A. B. Mhenni, D. V. Tuan, L. Geilen, M. M Petric, M. Erdi, K. Watanabe, T. Taniguchi, S. A. Tongay, K. Muller, N. P. Wilson, J. J. Finley, H. Dery, M. Barbone, 
Breakdown of the static dielectric screening approximation of Coulomb interactions in atomically thin semiconductors, ACS Nano \textbf{19}, 4269 (2025).  



\bibitem{Haug_SchmittRink_PQE84} H. Haug and S. Schmitt-Rink, Electron theory of the optical properties of laser excited semiconductors,  Prog. Quant. Electr. \textbf{9}, 3 (1984). %



   \bibitem{Forste_NatCom20}
J. Forste, N. V. Tepliakov, S. Y. Kruchinin, J. Lindlau, V. Funk, M. Forg, K. Watanabe, T. Taniguchi, A. S. Baimuratov, and A. Hogele, Exciton g-factors in monolayer, and bilayer WSe$_2$ from experiment, and theory, Nat. Commun. {\bf 11}, 4539 (2020).




\bibitem{Lindlau_NatComm18}
J. Lindlau, M. Selig, A. Neumann, L. Colombier, J. Forste, V. Funk, M. Forg, J. Kim, G. Berghauser, T. Taniguchi, K. Watanabe, F. Wang, E. Malic, and A. Hogele, The role of momentum-dark excitons in the elementary optical response of bilayer WSe$_2$, Nat.  Commun. {\bf 9}, 2586 (2018). 


\bibitem{Liu_PRR19}
E. Liu, J. v. Baren, T. Taniguchi, K. Watanabe, Y.-C. Chang, and C. H. Lui, Valley-selective chiral phonon replicas of dark excitons and trions
in monolayer WSe$_2$. Phys. Rev. Res. 1, 032007 (2019).

 \bibitem{VanTuan_PRB25}
 D. V. Tuan and H. Dery, Component exchange theory of trions,  Phys. Rev. B {\bf 111}, 085305 (2025). 

\bibitem{Rytova_MSU67}  N. S. Rytova, \textit{Screened potential of a point charge in a thin film},  Proc. MSU, Phys. Astron. 3, 30 (1967).
\bibitem{Keldysh_JETP79} L. V. Keldysh, \textit{Coulomb interaction in thin semiconductor and semimetal films}, JETP Lett. \textbf{29}, 658 (1979). %
%
\bibitem{Cudazzo_PRB2011} P. Cudazzo, I. V. Tokatly, and A. Rubio, Dielectric screening in two-dimensional insulators: Implications for excitonic and impurity states in graphene, Phys. Rev. B {\bf 84}, 085406 (2011).



\bibitem{VanTuan_PRB22}  D. V. Tuan and H. Dery,  Composite excitonic states in doped semiconductors, Phys. Rev. B \textbf{106}, L081301 (2022).

 \bibitem{VanTuan_ArXiv22}
 D. V. Tuan and H. Dery, Turning many-body problems to few-body ones in photoexcited semiconductors using the stochastic variational method in momentum space, SVM-k,  arXiv:2202.08378. 


\bibitem{VanTuan_PRX17} D. Van Tuan, B. Scharf, I. {\u Z}uti{\'c}, and H. Dery, Marrying excitons and plasmons in monolayer transition-metal dichalcogenides, Phys. Rev. X \textbf{7}, 041040 (2017).

\bibitem{HaugBook} H. Haug and S. W. Koch, \textit{Quantum theory of the optical and electronic properties of semiconductors}, 3rd ed. (World Scientific, Singapore, 1994).

\bibitem{Kormanyos_2DMater15}   A. Korm\'{a}nyos, G. Burkard, M. Gmitra, J. Fabian, V. Z\'{o}lyomi, N. D. Drummond, and V. Fal'ko, \textit{$\mathbf{k \cdot p}$ theory for two-dimensional transition metal dichalcogenide semiconductors}, 2D Mater. \textbf{2}, 022001 (2015). 



\bibitem{Cai_SSC2007}
Y. Cai, L. Zhang, Q. Zeng, L. Cheng, and Y. Xu, Infrared reflectance spectrum of BN calculated from first principles, Solid State Commun. {\bf 141},  262 (2007). 
%
%
\bibitem{Dai_Science2014}
S. Dai, Z. Fei, Q. Ma, A. S. Rodin, M. Wagner, A. S. McLeod, M. K. Liu, W. Gannett, W. Regan, K. Watanabe, T. Taniguchi, M. Thiemens, G. Dominguez, A. H. Castro Neto, A. Zettl, F. Keilmann, P. Jarillo-Herrero, M. M. Fogler, and D. N. Basov, Tunable phonon polaritons in atomically thin van der Waals crystals of boron nitride, Science {\bf  343}, 1125 (2014).
%

\bibitem{Stier_PRL18} A. V. Stier, N. P. Wilson, K. A. Velizhanin, J. Kono, X. Xu, and S. A. Crooker, Magnetooptics of exciton Rydberg states in a monolayer semiconductor, Phys. Rev. Lett. \textbf{120}, 057405 (2018).

\end{thebibliography}
\end{document}